\documentclass[twocolumn,reprint,epsf,graphics,psfig,superscriptaddress]{revtex4}

\usepackage{amsfonts}
\usepackage{graphicx}
\usepackage{amsmath}
\usepackage{amssymb}
\usepackage{latexsym}
\usepackage{psfrag}
\usepackage{dcolumn}% Align table columns on decimal point
\usepackage{datetime}
\usepackage{multirow}
\usepackage{subfigure}
\usepackage{amssymb,amsmath,rotate,color}

\begin{document}
\title{Malleability at the extreme nanoscale: Slow and fast
quakes of few-body systems}
\author{Alireza Saffarzadeh}
\altaffiliation{Author to whom correspondence should be addressed. Electronic mail: asaffarz@sfu.ca}
\affiliation{Department of Physics, Payame Noor University, P.O.
Box 19395-3697 Tehran, Iran} \affiliation{Department of Physics,
Simon Fraser University, Burnaby, British Columbia, Canada V5A
1S6}
\author{George Kirczenow}
\affiliation{Department of Physics, Simon Fraser University,
Burnaby, British Columbia, Canada V5A 1S6}
\date{\today}

\begin{abstract}
We explore the malleability of ultra-small metal nanoparticles by
means of {\em ab initio} calculations. It is revealed that, when
strained, such nanoparticles exhibit complex behavior, including
bifurcation between slow and fast quakes of their atomic
structure, despite being few-body systems. We show the bifurcation
to arise from the collapse of the nanoparticle's stiffness and a
broken soft mode symmetry, and that whether a slow or fast quake
occurs can be controlled by varying the amplitude of the
externally applied strains. We predict that while energy is
released abruptly in a fast quake, surprisingly, it continues to
build up during a slow quake and that, in common with slow-slip
geological earthquakes, the slow nanoparticle quake is a silent
precursor to a fast ``seismic" quake. We show that electrical
conductance and force measurements can detect and distinguish
between slow and fast quakes, opening the way for experiments and
potential applications of these phenomena.
\end{abstract}
\maketitle
%{\bf PACS.}

\section{Introduction}
The malleability of metals, i.e., their ability to deform
plastically without fracturing, has been exploited since
prehistoric times in the making of ornaments, tools and weapons.
It continues to have numerous practical applications today, and
the study of plasticity, shear and fracture of macroscopic matter
is an active area of
research.\cite{Chen17,Denisov,Baros14,Baros13,Grob,
Bonamy,Langer,Maloy,Falk98,Falk08} However, little, if anything,
is known at present, either experimentally or theoretically,
regarding the malleability of nanoparticles, despite its potential
importance and the long-standing, intense interest in the physics
of nanoparticles, including their
structural,\cite{Johnson,Hock,Chen}
electronic,\cite{Ralph,Agam,Lykhach} optical,\cite{Zheng,Barbry}
plasmonic,\cite{Scholl,Barbry} magnetic,\cite{Deshmukh,Hai}
thermal \cite{Hock} and transport\cite{Ralph,Agam,Deshmukh}
properties. Here we explore the malleability of ultra-small ($<$1
nm) metal nanoparticles theoretically by means of first principles
computer simulations of the deformation of copper atomic clusters,
as may be realized in a mechanically controlled break junction or
scanning tunneling microscope setup. We establish that these
nanoparticles exhibit malleability as distinct from brittleness.
We find that under uniaxial elongation or compression, the
nanoparticle normally deforms gradually, at first elastically, and
then plastically. However, these gradual deformations are
punctuated by sudden events that we will refer to as ``quakes."
During quakes some of the interatomic separations switch rapidly
between values typical of nearest neighbor and second neighbor
distances. Given the seeming simplicity of the few atom
nanoparticles that we consider, the behavior of the quakes is
revealed by the present investigation to be surprisingly complex
and interesting. We find quakes of two different types: (i) Fast
quakes in which the atomic configuration changes abruptly and the
total energy of the nanoparticle {\em decreases discontinuously},
and (ii) Slow quakes in which the atomic configuration changes in
a series of closely spaced smaller steps while, unexpectedly, the
total energy {\em increases} smoothly. We predict bifurcation to
occur between slow and fast quakes due to the collapse of the
nanoparticle's stiffness and breaking of a symmetry associated
with soft vibrational modes. We demonstrate that we are able to
{\em control} whether a slow or fast quake occurs by choosing the
{\em amplitude} of the forcing uniaxial strain applied to the
nanoparticle. We find that the slow and fast quakes result in
differing atomic configurations of the nanoparticle, and that a
slow quake acts as a {\em precursor} to a subsequent fast quake
that results in yet a third atomic configuration. We also show
that conductance and force measurements can distinguish between
fast and slow quakes. The behavior of the slow and fast quakes
that we have discovered in few-atom nanoparticles has surprising
commonalities with geological subduction zone earthquakes. Namely,
in common with our findings for nanoparticle quakes, it has been
suggested that some violent (fast) earthquakes follow silent
slow-slip precursor earthquakes,\cite{Schwartz, Socquet} and that
whether a geological fault alternates between slipping slowly
(silently) and fast (seismically) depends on the loading
conditions.\cite{Mclaskey} However, in contrast to a recent
proposal regarding geological earthquakes, \cite{Mclaskey} we find
the reduced stiffness signaled by the onset of a soft mode to play
a crucial role in slow nanoparticle quakes.

\section{Methodology}
In the remainder of this paper we elucidate these remarkable
findings by presenting representative results for nanoparticles
consisting of eight copper atoms. The atomic positions and
interatomic distances were estimated by minimizing the total
energy of the system, computed within density functional theory
using the GAUSSIAN 09 package with the PBE1PBE functional and
Lanl2DZ pseudopotentials and basis sets \cite{Frisch,Perdew}. In
all optimized geometries, the electronic energy and ionic forces
were converged within $10^{-5}$ eV and 0.0008 eV/{\AA},
respectively.

Initially, the structure was optimized without any constraint to
find the minimum energy $E_0$, the corresponding arrangement of
the copper atoms, and the distances $d_{ij}$ between atoms $i$ and
$j$, $d_{ij}=|\mathbf{R}_{j}-\mathbf{R}_{i}|$. Here
$\mathbf{R}_{i}$ is the position of atom $i$. The lowest energy
structure (shown in inset (a) of Fig.\ref{F1}) was found to have
tetrahedral symmetry, i.e., the point group $T_d$. Its first,
second and third neighbor atomic distances were found to be 2.48,
4.08 and 4.12{\AA}, respectively.

Starting from this minimum energy structure, the nanoparticle was
strained as follows: A pair of atoms $i$ and $j$ was chosen and
their separation $d_{ij}$ was increased by 0.05{\AA}. Then atoms
$i$ and $j$ were frozen and the other atoms of the nanoparticle
were allowed to relax so as to minimize the total energy. This
elongation of $d_{ij}$ and energy optimization were repeated
multiple times. Our results, obtained by increasing the distance
between atoms 1 and 2, $d_{12}$, stepwise in this way are
presented in Fig.\ref{F1}. Qualitatively similar behavior was
found if other atoms were chosen instead of 1 and 2.

\section{Results and discussion}
\begin{figure}
\centerline{\includegraphics[width=0.85\linewidth]{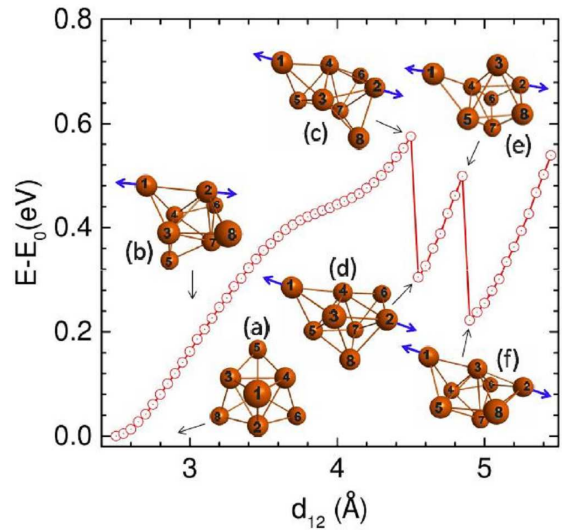}}
\caption{(Color online) Variation of the total energy of the
eight-atom copper cluster during stretching of the interatomic
distance $d_{12}$. Inset (a): The lowest energy geometry of the
unconstrained cluster. (b)-(f): Atomic arrangements of the cluster
for different stretching lengths $d_{12}$ \cite{Mac}. Blue arrows
mark atoms 1 and 2 and show the direction of elongation for each
structure. Black arrows indicate the corresponding $d_{12}$
values. The lines are guides to the eye. Energies are measured
relative to the lowest energy $E_0$ of the unstrained
nanoparticle.} \label{F1}
\end{figure}

Fig.\ref{F1} shows how the total energy of the nanoparticle
changes as the cluster is stretched in this way, and also how the
nanoparticle's geometry evolves in this process. Stretching the
copper cluster initially increases the total energy in a
parabola-like way due to the quadratic effective interatomic
potential for small atomic displacements from their equilibrium
positions. In this regime Hooke's law is obeyed and thus the
deformation is elastic. However as the cluster is stretched
further from configuration (a) to configuration (b) the behavior
evolves continuously from elastic to plastic deformation, as the
total energy characteristic changes from parabolic for small
displacements (elastic deformation) to approximately linear
behavior (plastic deformation). In the plastic regime the bond
between atoms 1 and 2 slowly ruptures and this manifests itself as
an inflection point in the energy. As the deformation increases
further another inflection point appears, due to a new bond
gradually forming between atoms 1 and 5 as the cluster evolves
from structure (b) to (c).

Although plastic deformation of the cluster due to changes in the
bonding topology occurs as $d_{12}$ is stretched between
configurations (a) and (c) in Fig.\ref{F1}, both the total energy
and all of the interatomic distances evolve smoothly and slowly
throughout this process. In other words, substantial changes in
these quantities occur gradually as $d_{12}$ changes by $\sim
1${\AA}. By contrast the transitions between configurations (c)
and (d) and between (e) and (f) in Fig.\ref{F1} involve abrupt
decreases in the total energy due to the sudden formation of new
bonds. These events are examples of `fast quakes'. They arise from
instabilities built up in the system. In the transition from (c)
to (d) a new bond forms abruptly between atoms 8 and 5, increasing
the coordination numbers of atoms 8 and 5 and thus lowering the
total energy. Similarly, in the transition from (e) to (f) new
bonds form between atom 3 and atoms 1, 6 and 7, lowering the total
energy abruptly. These results show that metal nanoparticles with
dimensions as small as two times the nearest neighbor distance can
exhibit malleable behavior as distinct from brittle fracture.
\begin{figure}
\centerline{\includegraphics[width=1.0\linewidth]{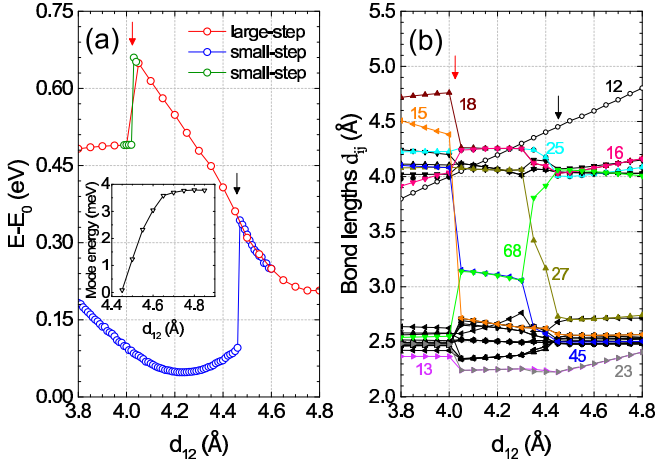}}
\caption{(Color online) (a) Total energy of the copper cluster
during compression of $d_{12}$ by steps of 0.01{\AA} (blue and
green circles) and 0.05{\AA} (red circles). Inset: Energy of the
soft vibrational mode approaches zero for $d_{12}\sim4.45${\AA}
where both fast and slow quakes begin. (b) Bond lengths $d_{ij}$
(labelled ${ij}$) plotted vs $d_{12}$ for 0.05{\AA} compression
steps. Most $d_{12}$ values cluster around first and second
neighbor distances. In both (a) and (b) red (black) arrow
indicates where only fast (both fast and slow) quake(s) occur.
Lines are guides to the eye.} \label{F2}
\end{figure}

In Fig.\ref{F2} we show the behavior of the nanoparticle if after
stretching it (as in Fig.\ref{F1}) we compress it by reducing the
separation $d_{12}$ in similar steps, again freezing the positions
of atoms 1 and 2 and allowing the other atoms to relax after each
step. The starting geometry for compressing the cluster is the
stretched structure with $d_{12}$=5.45{\AA}. In Fig.\ref{F2}(a)
the total energy of the cluster is shown in red, blue and green
for $d_{12}$ compressed in larger and smaller steps, 0.05{\AA},
0.01{\AA} and 0.01{\AA}, respectively.

Comparing Fig.\ref{F2}(a) with Fig.\ref{F1} it is evident that the
deformations of the cluster are not reversible. In particular, for
elongation of $d_{12}$ in 0.05{\AA} steps the cluster exhibits 2
fast quakes, at $d_{12}$=4.55{\AA} and $d_{12}$=4.9{\AA} in
Fig.\ref{F1}. By contrast, under compression,  for the 0.05{\AA}
$d_{12}$ step size [red circles in Fig.\ref{F2}(a)] there is
instead one fast quake at $d_{12}$=4{\AA} indicated by the red
arrow.

However, surprisingly, for compression of $d_{12}$ by smaller
0.01{\AA} steps [blue circles in Fig. \ref{F2} (a)] a fast quake
appears in the total energy at $d_{12}=4.45${\AA} indicated by the
black arrow, whereas no feature is evident in the energy for
$d_{12}$ values in that vicinity for the larger 0.05{\AA} $d_{12}$
compression steps [red circles in Fig.\ref{F2} (a)].

To help understand this intriguing anomaly we plot all of the
interatomic distances  $d_{ij}$ of the nanoparticle against
$d_{12}$ for the larger 0.05{\AA} $d_{12}$ compression steps in
Fig.\ref{F2} (b). Notice that, at the same value of
$d_{12}=4.45${\AA} (indicated by the black arrow) where the
`anomalous' fast quake for small steps occurs in Fig.\ref{F2}(a),
in Fig.\ref{F2}(b) the values of  $d_{27}$ and $d_{68}$ begin to
change. After three 0.05{\AA} $d_{12}$ steps, $d_{27}$ has
switched from a first neighbor value near 2.7{\AA} to a second
neighbor value near 4{\AA} whereas $d_{68}$ has switched by a
comparable amount in the opposite direction. Because these large
(1{\AA} or greater) changes in the interparticle distances
($d_{27}$ and $d_{68}$) occur during a small (0.15{\AA}) change in
the forcing strain applied to $d_{12}$ this behavior constitutes a
quake. But in contrast to fast quakes that occur within a single
$d_{12}$ step, this quake is spread over multiple steps. We
therefore refer to it as a ``slow quake". Furthermore,
interestingly, unlike for fast quakes in which the total energy
{\em drops} abruptly, in the slow quake the total energy continues
to {\em rise} almost linearly while $d_{12}$ changes, as is shown
by the red circles near the black arrow in  Fig.\ref{F2}(a).

It is not a coincidence that this slow quake in Fig.\ref{F2}(b)
and the fast quake marked by the black arrow in Fig.\ref{F2}(a),
{\em both} begin at the same value 4.45{\AA} of $d_{12}$. It
points to interesting physics that will be explained below.

\begin{figure}
\centerline{\includegraphics[width=0.95\linewidth]{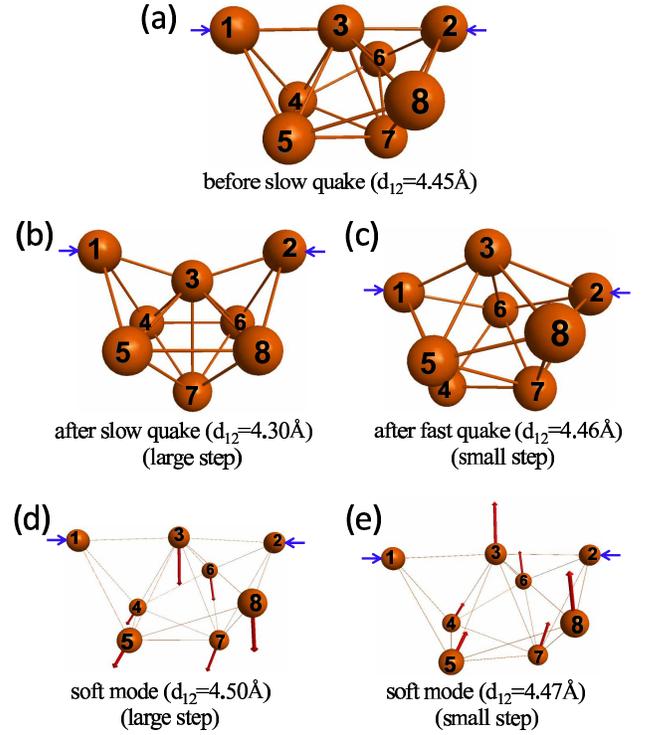}}
\caption{(Color online) Atomic arrangements of the copper
nanoparticle \cite{Mac} for compression along $d_{12}$ (a) before
and (b) after slow quake, and (c) after fast quake. (d) and (e):
Atomic arrangements and soft mode (red vectors) during compression
by large and small steps, respectively. The blue arrows show the
direction of $d_{12}$ compression in each structure.} \label{F3}
\end{figure}

Fig.\ref{F3}(a) shows the structure of the nanoparticle just
before these slow and fast quakes begin, while the structures
immediately after the slow and fast quakes are shown in
Figs.\ref{F3}(b) and (c) respectively. Comparing these pictures,
it is clear that in the slow and fast quakes the atoms of the
nanoparticles are displaced in opposite directions. For example,
after the slow quake [Fig.\ref{F3}(b)] atom 3 is lower than in the
pre-quake structure [Fig.\ref{F3}(a)] whereas after the fast quake
[Fig.\ref{F3}(c)] the same atom 3 is higher. In each case the
other atoms of the nanoparticle (apart from the forcing atoms 1
and 2) have moved in roughly the same direction as atom 3.

The physical reason for this symmetry between the atomic
displacements of the slow and fast quakes can be understood by
considering the vibrational normal modes of the copper
nanoparticle with atoms 1 and 2 frozen. The calculated vibron
energy for one of those modes is shown as a function of $d_{12}$
in the inset of Fig.\ref{F2}(a). It is evident that this is a soft
mode whose frequency approaches zero near $d_{12}=$4.45{\AA} where
the  slow and fast quakes both begin. The normalized atomic
displacements in the soft mode are shown as red arrows in
Fig.\ref{F3}(d) and (e) for the nanoparticle geometries
immediately preceding the slow and fast quakes. The atomic
displacements shown in Fig.\ref{F3}(d) and (e) clearly belong to
the same soft mode; they differ only in that they are
phase-shifted by $\pi$ relative to each other.

Most importantly, the atomic displacements that happen in the slow
and fast quakes [Fig.\ref{F3}(b) and (c)] are strikingly similar
to the soft mode atomic displacements shown in  Fig.\ref{F3}(d)
and (e), respectively. We conclude that these slow and fast quakes
both occur because the stiffness of the nanoparticle against the
soft mode-like atomic displacements [shown in Fig.\ref{F3}(d) and
(e)] collapses when the soft mode frequency approaches zero for
$d_{12}\sim4.45${\AA}. In the harmonic approximation {\em small}
atomic displacements in the opposite directions shown in
Fig.\ref{F3}(d) and (e) require equal amounts of energy. However
for the large displacements that occur in the fast and slow quakes
this symmetry is broken by anharmonicity. In addition, the slow
quake type displacements [Fig.\ref{F3}(b) and (d)] cost additional
energy whereas for fast quake type displacements [Fig.\ref{F3}(c)
and (e)] the total energy of the nanoparticle decreases
spontaneously. This difference in the energetics between the fast
and slow quakes is evident from the different behavior of the red
(slow quake) and blue (fast quake) energy plots in Fig.\ref{F2}(a)
as  $d_{12}$ decreases below 4.45{\AA} as marked by the black
arrow. Compression of $d_{12}$ by larger (smaller) 0.05{\AA}
(0.01{\AA}) steps supplies more (less) energy to the nanoparticle.
Thus the larger compression steps result in the slow quake in
which the total energy increases, while the smaller steps result
in the fast quake in which the total energy decreases. Thus we can
{\em control} whether a slow or fast quake happens by choosing the
amplitude of the forcing strains applied to $d_{12}$. In other
words, this demonstrates that if we squeeze a nanostructure
gradually, we can get a very different result than by applying
larger abrupt strains akin to hammering it. Both cases are
physical, with different ways of compressing the structure
producing different results. Here we have assumed that the abrupt
compression of $d_{12}$ is fast compared to the equilibration time
of the nanoparticle structure. It is reasonable to expect the
equilibration to become slower at lower temperatures and also as
the soft mode frequency approaches zero.

Another important feature of Fig.\ref{F2}(a) is that if the
nanoparticle continues to be compressed after the slow quake has
occurred the total energy of the system continues to build almost
linearly until the fast quake indicated by the red arrow occurs.
I.e., the slow quake acts as a precursor to this fast quake.
[Reducing the step size does not alter the nature of this
subsequent fast quake, as can be seen by comparing the red and
green circles in Fig.\ref{F2}(a).] We note that since there is no
release of energy during a slow quake (the total energy continues
to grow during the slow quake) the slow quake should be regarded
as ``silent" as is the case for slow geological
earthquakes.\cite{Schwartz,Socquet,Mclaskey} It should also be
noted that we found qualitatively similar behavior of fast and
slow quakes and their relationships as well as similar underlying
physics if instead of $d_{12}$ we stretched and compressed other
interparticle distances  $d_{ij}$ in similar ways.
\begin{figure}
\centerline{\includegraphics[width=0.95\linewidth]{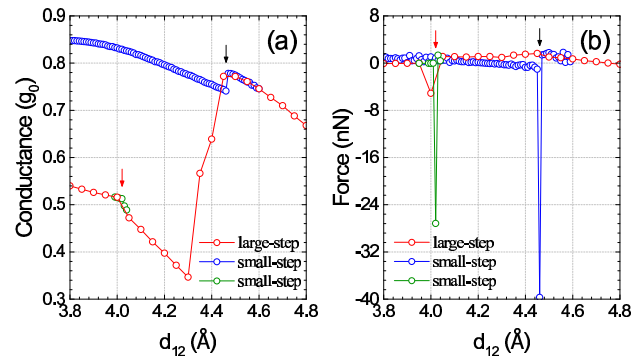}}
\caption{(Color online) (a) Calculated conductance and (b) the
mechanical force required to compress the cluster when the
distance $d_{12}$ is compressed by small (blue and green circles) and
large (red circles) steps. The arrows indicate
the $d_{12}$ values at which the slow and fast quakes appear;
their colors and locations are as in Fig.\ref{F2}.
The lines are guides to the eye.} \label{F4}
\end{figure}

Since quakes alter the structure of the nanoparticle,
the total energy and related, potentially experimentally
accessible, physical properties such as electronic conductance and the applied
force required for a mechanical deformation can be affected strongly.

To investigate the effect of quakes on electronic transport
through the nanoparticle, we calculated the electron transmission
probabilities $T$ through the cluster at the Fermi energy
$\epsilon_F$ of the macroscopic electrodes used in conductance
measurements. These, within Landauer theory,\cite{Datta} are
related to  the conductance via
$g(\epsilon_F)=g_{0}T(\epsilon_F)$, where $g_{0}=2e^2/h$ is the
conductance quantum. Here,
$T=\sum_{\alpha,\beta}|t_{\beta\alpha}|^2\frac{v_\beta}{v_\alpha}$,
$t_{\beta\alpha}$ is the transmission amplitude through the
cluster, and $\alpha$ ($\beta$) is the electronic state of a
carrier with velocity $v_\alpha$ ($v_\beta$) in the source (drain)
lead. The leads are attached to the two copper atoms that are
under compression, with separation $d_{ij}$. Fig.\ref{F4}(a) shows
how the conductance at the Fermi energy, $g(\epsilon_F)$, evolves
during compression by small and large steps, vs. parameter
$d_{12}$. The slow quake results in a large (factor $\sim$2)
smooth decrease in conductance, while fast quakes appear as
abrupt, small conductance steps. The large conductance decrease
during the slow quake can be understood intuitively as a result of
the less compact structure after the slow quake [compare Fig.
\ref{F3}(b) with \ref{F3}(a) and (c)] which also accounts for the
{\em increase} in the total energy of the nanoparticle during the
slow quake in Fig. \ref{F2}(a).

The mechanical force, $F_m$, during compression of the copper
cluster can be calculated from the numerical derivative of the
total energy with respect to the length parameter $d_{12}$ using
the equation:
$F_m(d_{12,i})=-[E(d_{12,i+1})-E(d_{12,i})]/\delta_i$ where
$d_{12,i}$ is the length parameter of compression step $i$,
$E(d_{12,i})$ is the total energy in terms of $d_{12,i}$, and
$\delta_i=d_{12,i+1}-d_{12,i}$. We have depicted in Fig.
\ref{F4}(b) the magnitude of the force versus parameter $d_{12}$
for small-step and large-step compressions. The force shows a
strong spike whenever a fast quake occurs because the total energy
decreases abruptly, but no comparable feature is displayed for a
slow quake.

Therefore, as can be seen in Fig. \ref{F4}(a) and (b), the
conductance and force together can in principle distinguish
between slow and fast quakes experimentally. Hence, in addition to
being of fundamental interest, experimental studies of
malleability at the nanoscale based on these findings may be
relevant for nanoelectronic applications.

\section{Conclusions}
In summary, our investigation has established theoretically that
ultra-small metal nanoparticles are malleable, and has revealed
surprising bifurcatory behavior of the response of nanoparticles
to externally applied strains. The bifurcation arises from
anharmonicity breaking the inversion symmetry of a soft
vibrational mode of the nanoparticle. It manifests as slow and
fast quakes that alter the structure of the nanoparticle. We have
shown that whether a slow or fast quake occurs can be controlled
by varying the amplitude of the applied strain. We have also shown
that energy is released abruptly in a fast quake but, by contrast,
it continues to build up during a slow quake. In common with
slow-slip geological earthquakes, we predict that the slow
nanoparticle quake will be a silent precursor of a fast
``seismic'' quake. We have shown that electrical conductance and
force measurements can detect and distinguish between slow and
fast quakes, opening the way to experiments and potential
applications of this interesting phenomenon.

\section*{Acknowledgements}
This work was supported by NSERC, CIFAR, WestGrid, and Compute
Canada.

\end{document}